# The Evolution of Gene Dominance through the Baldwin Effect

Larry Bull

Computer Science Research Centre

Department of Computer Science & Creative Technologies

University of the West of England, Bristol UK

Larry.Bull@uwe.ac.uk

**Abstract**

It has recently been suggested that the fundamental haploid-diploid cycle of eukaryotic sex exploits a rudimentary form of the Baldwin effect. Thereafter the other associated phenomena can be explained as evolution tuning the amount and frequency of learning experienced by an organism. Using the well-known NK model of fitness landscapes it is here shown that the emergence of dominance can also be explained under this view of eukaryotic evolution.

Keywords: Diploid, Eukaryotes, Meiosis, NK model, Recombination

**Introduction**

Dominance has typically been explained either as a consequence of enzymatic pathways with selection playing little or no role (eg, [Wright, 1929]) or as a consequence of maintained periods of high degrees of allele heterogeneity (eg, [Clark, 1964]). Using a recently presented view of eukaryotic evolution, a new explanation for the emergence of dominance can be proposed. Eukaryotic sex is here defined as successive rounds of syngamy and meiosis in a haploid-diploid lifecycle. It has been suggested that the emergence of a haploid-diploid cycle enabled the exploitation of a rudimentary form of the Baldwin effect [Baldwin, 1896] and that this provides an underpinning explanation for all the observed forms of sex [Bull, 2017]. The Baldwin effect is here defined as the existence of phenotypic plasticity that enables an organism to exhibit a significantly different (better) fitness than its genome directly represents. Over time, as evolution is guided towards such regions under selection, higher fitness alleles/genomes which rely less upon the phenotypic plasticity can be discovered and become assimilated into the population.

Key to the new explanation for the evolution of sex in eukaryotes is to view the process from the perspective of the constituent haploids. A diploid organism may been seen to simultaneously represent two points in the underlying haploid fitness landscape. The fitness associated with those two haploids is therefore the fitness achieved in their combined form as a diploid; each haploid genome will have the same fitness value and that will almost certainly differ from that of their corresponding haploid organism due to the interactions between the two genomes. That is, the effects of haploid genome combination into a diploid can be seen as a simple form of phenotypic plasticity for the individual haploids before they revert to a solitary state during reproduction. In this way evolution can be seen to be both assigning a single fitness value to the *region* of the landscape between the two points represented by a diploid's constituent haploid genomes and altering the shape of the haploid fitness landscape. In particular, the latter enables the landscape to be smoothed under a rudimentary Baldwin effect process [Hinton & Nowlan, 1987].

Numerous explanations exist for the benefits of recombination (eg, [Bernstein and Bernstein, 2010]) but the role becomes clear under the new view: recombination facilitates genetic assimilation within the simple form of the Baldwin effect. If the haploid pairing is beneficial and the diploid is chosen under selection to reproduce, the recombination process can bring an assortment of those partnered genes together into new haploid genomes. In this way the fitter allele values from the pair of partnered haploids may come to exist within individual haploids more quickly than the under mutation alone (see [Bull, 2017] for full details).

In this short paper, following [Bull, 2017], the evolution of gene dominance is explored using versions of the well-known NK model [Kauffman & Levin, 1987] of fitness landscapes where size and ruggedness can be systematically altered. Results suggest that varying degrees of dominance can be shown to be beneficial under various conditions as it enables variation in the amount of learning occurring on rugged fitness landscapes.

**The NK Model**

Kauffman and Levin [1987] introduced the NK model to allow the systematic study of various aspects of fitness landscapes (see [Kauffman, 1993] for an overview). In the standard model, the features of the fitness landscapes are specified by two parameters: $N$, the length of the genome; and $K$, the number of genes that has an effect on the fitness contribution of each (binary) gene. Thus increasing $K$ with respect to $N$ increases the epistatic linkage, increasing the ruggedness of the fitness landscape. The increase in epistasis increases the number of optima, increases the steepness of their sides, and decreases their correlation. The model assumes all intragenome interactions are so complex that it is only appropriate to assign random values to their effects on fitness. Therefore for each of the possible $K$ interactions a table of $2^{(K+1)}$ fitnesses is created for each gene with all entries in the range 0.0 to 1.0, such that there is one fitness for each combination of traits (Figure 1). The fitness contribution of each gene is found from its table. These fitnesses are then summed and normalized by $N$ to give the selective fitness of the total genome.

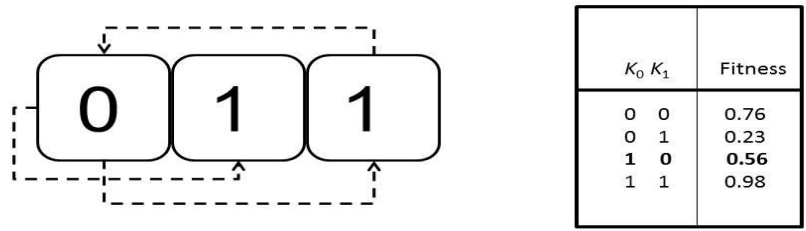

Figure 1: An example NK model ($N=3$, $K=1$) showing how the fitness contribution of each gene depends on $K$ random genes (left). Therefore there are $2^{(K+1)}$ possible allele combinations per gene, each of which is assigned a random fitness. Each gene of the genome has such a table created for it (right, centre gene shown). Total fitness is the normalized sum of these values.

Kauffman [1993] used a mutation-based hill-climbing algorithm, where the single point in the fitness space is said to represent a converged species, to examine the properties and evolutionary dynamics of the NK model. That is, the population is of size one and a species evolves by making a random change to one randomly chosen gene per generation. The "population" is said to move to the genetic configuration of the mutated individual if its fitness is greater than the fitness of the current individual; the rate of supply of mutants is seen as slow compared to the actions of selection. Ties are broken at random. Figure 2 (top row) shows example results. All results reported in this paper are the average of 10 runs (random start points) on each of 10 NK functions, ie, 100 runs, for 20,000 generations. Here $0 \leq K \leq 15$, for $N=20$ and $N=100$.

As discussed in [Maynard-Smith & Szathmary, 1995, p150], the first step in the evolution of eukaryotic sex was the emergence of a haploid-diploid cycle, probably via endomitosis, then simple syngamy or one-step meiosis, before two-meiosis with recombination . Following [Bull, 2017], the NK model can be extended to consider aspects of the evolution of sexual diploids. Firstly, each individual contains two haploid genomes of the form described above for the standard model. The fitness of an individual is here simply assigned as the average of the fitness of each of its constituent haploids. These are initially created at random, as before. Two-step meiosis with recombination is here implemented as follows: on each generation the diploid individual representing the converged population is copied twice to create two offspring. In each offspring, each haploid genome is copied once, a single recombination point is chosen at random, and non-sister haploids are recombined. One of the four resulting haploids in each offspring individual is chosen at random. Finally, a random gene in each chosen haploid is mutated. The resulting pair of haploids forms the new diploid offspring to be evaluated.

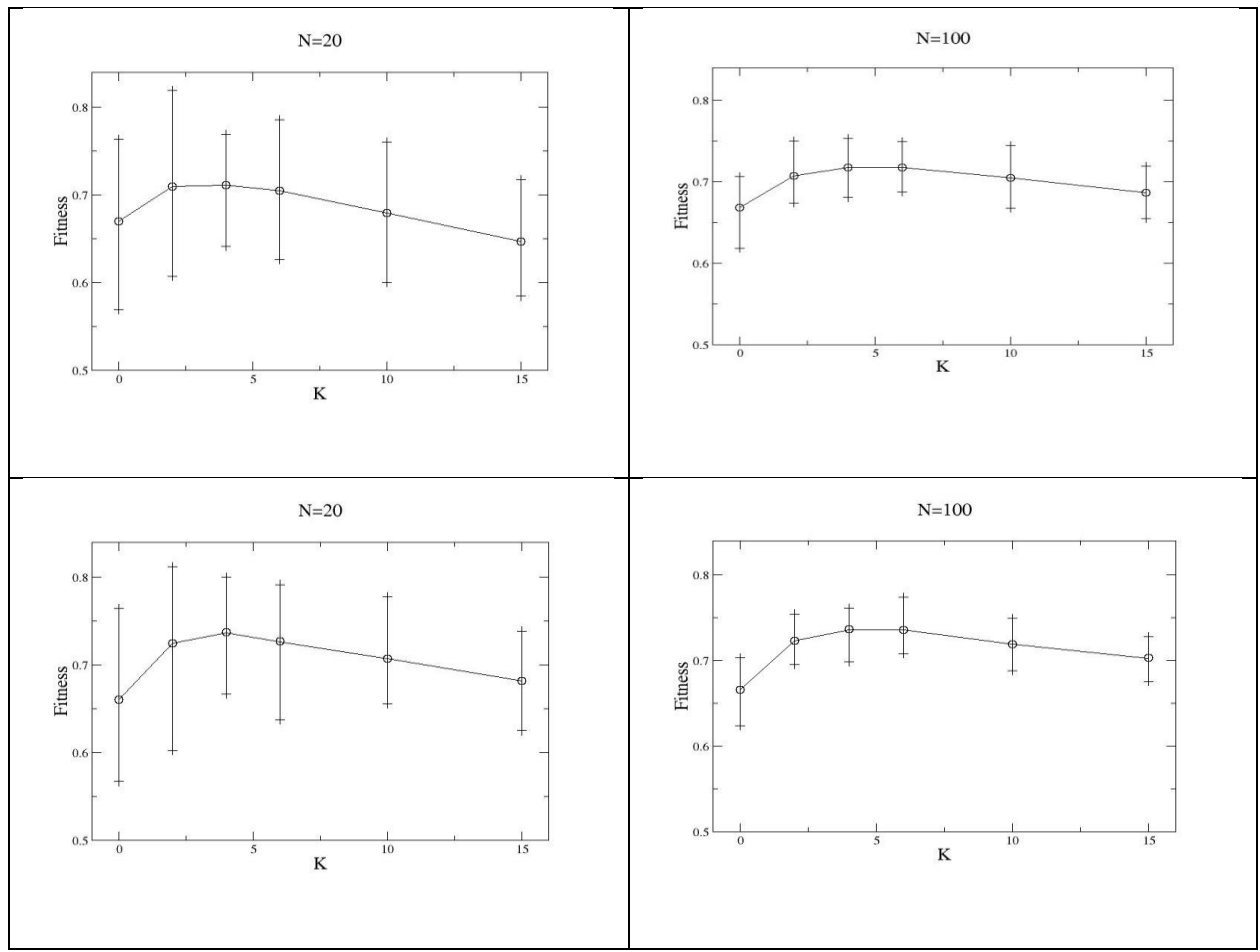

Figure 2: Showing typical behaviour and the fitness reached after 20,000 generations on landscapes of varying ruggedness ($K$) and length ($N$) for asexual haploids (top row) and sexual diploids (bottom row) Error bars show min and max values.

Figure 2 (bottom row) shows examples of the general properties of adaptation in the NK model of rugged fitness landscapes for diploid organisms evolving via two-step meiosis with recombination. When $N$=20 the fitness level reached is significantly lower than for $N$=100 for $K$>4 (T-test, $p$<0.05), as is seen in the traditional haploid case due to the effects of the increased landscape complexity. Following [Bull, 2017], it can be seen that fitness levels are always higher than the equivalent haploid case (top row) when $K$>0 due to the Baldwin effect as discussed (T-test, $p$<0.05).

**Dominance**

It was suggested in [Bull, 2017] that dominance can be explained as part of the Baldwin effect view of the evolution of eukaryotes since it can tune the amount of learning experienced on a per-gene basis. That is, dominance can be seen as a mechanism through which evolution is able to bias the composite fitness value assigned to the generalization over the region of the fitness landscape defined by the two constituent haploid genome end points. Hence it can be expected that the fraction of genes for which a dominant allele exists will vary with the ruggedness of the underlying fitness landscape.

The basic model above of two-step meiosis with recombination in a diploid can be extended to consider a simple dominance mechanism. Here an extra template of length $N$ is added to an individual where each locus can take one of three values – 0,1,#. The first two equate to the dominant allele value for that locus in the case of the two genomes being heterozygote and the last is said to indicate a lack of dominance and hence the fitness contribution is the average of the two genes, as used above. After meiosis, the offspring also has the dominance value of one randomly chosen locus in the template altered to another value, much like the standard gene mutation process. In the case of fitness ties between the offspring and parent, the one with fewer dominated alleles (most #s) is chosen, with subsequent ties broken at random as before. In this way, there is a slight selective bias against dominance. All other details remain as before and individuals are initialised without dominance, ie, $N$ #s in the template.

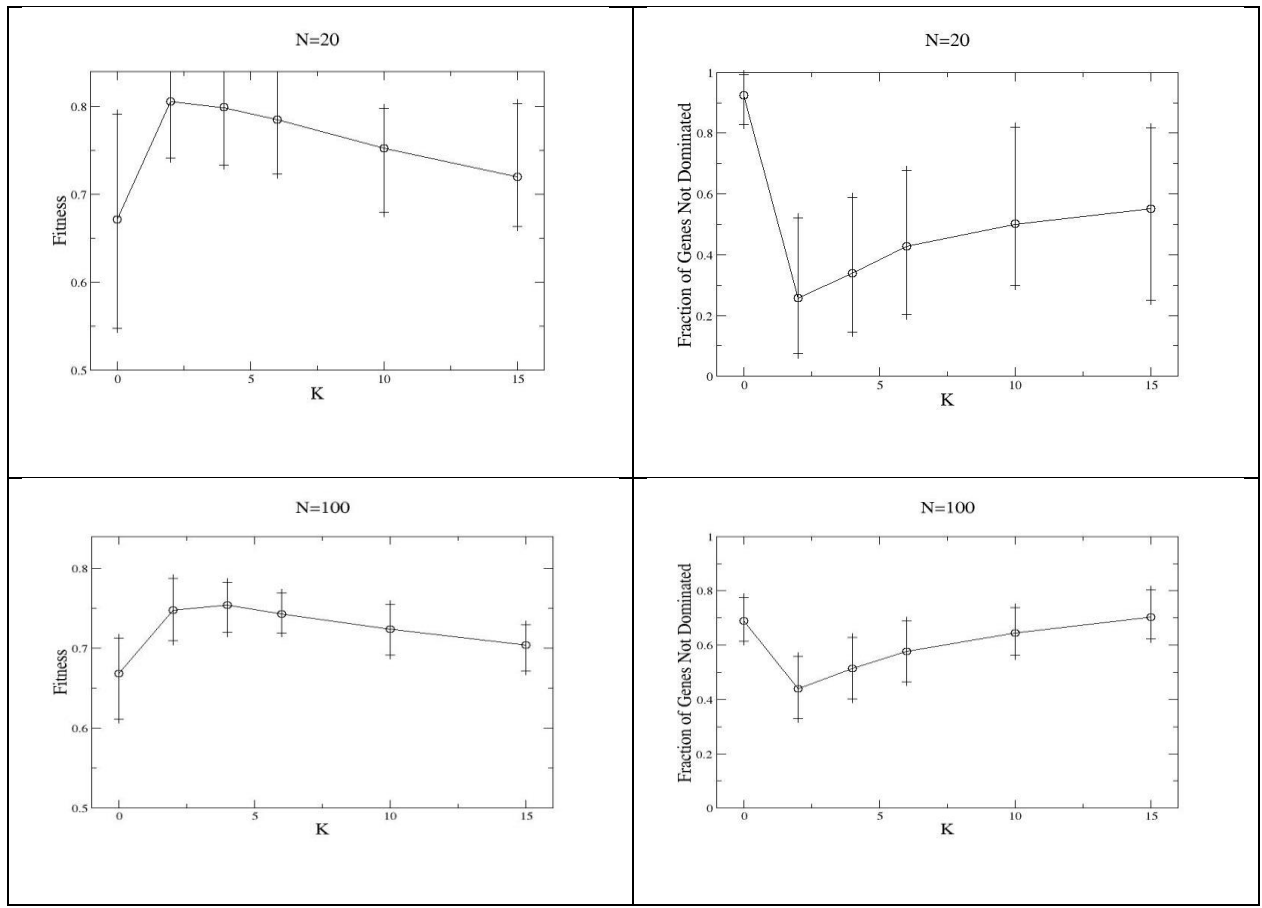

Figure 3: Showing typical fitness (left) reached after 20,000 generations on landscapes of varying ruggedness and size for two-step meiosis with recombination including a dominance mechanism (right).

Figure 3 shows how with $N$=20 fitness is significantly increased for $K$>0 and with $N$=100 fitness is increased for 0<$K$<15 (T-test, $p$<0.05, compare to Figure 4 right column). The biggest benefit is seen at $K$=2 in both cases where the most use of dominated genes emerges, dominance reducing with increasing $K$ thereafter. Hence the fitness bias mechanism for generalizations appears more useful on more correlated fitness landscapes not least since this is where the smoothing through the Baldwin effect is typically less beneficial (after [Bull, 1999]), apart from in the unimodal case of $K$=0. It can be noted that this general result is also seen when the genomes are initialised as homozygotes (not shown).

## Conclusion

"Periodic disagreements notwithstanding, the overall trend has been to explain the manifestation of dominance as a default—and invariant—property of biological systems" [Bagheri, 2006]. This paper has drawn upon a new explanation for the processes by which (sexual) eukaryotic organisms evolve to identify the selective advantage of dominance. Following [Bull, 2017], it has been shown to be an effective mechanisms through which evolution can tune the amount of fitness landscape smoothing experienced on a per-gene basis.